\documentclass[aps,prl,twocolumn,superscriptaddress,amsmath,amssymb]{revtex4-2}
\usepackage{graphicx}   
\usepackage{xcolor}     
\usepackage{hyperref}   
\usepackage{parskip}    
\usepackage{siunitx}
\usepackage{braket}
\newcommand{\ketbra}[1]{\ket{#1}\bra{#1} }

\begin{document}

\title{Surface-Selective Probe of Spin-Triplet Superconductivity in Rhombohedral Graphene}

\author{Kilian Krötzsch}
\thanks{These authors contributed equally}
\affiliation{Institute of Physics, École Polytechnique Fédérale de Lausanne (EPFL), CH-1015 Lausanne, Switzerland}

\author{Zekang Zhou}
\thanks{These authors contributed equally}
\affiliation{Institute of Physics, École Polytechnique Fédérale de Lausanne (EPFL), CH-1015 Lausanne, Switzerland}

\author{Kryštof Kolář}
\affiliation{Department of Applied Physics, Aalto University School of Science, FI-00076 Aalto, Finland}

\author{Yonggen Li}
\affiliation{Institute of Physics, École Polytechnique Fédérale de Lausanne (EPFL), CH-1015 Lausanne, Switzerland}

\author{Kenji Watanabe}
\affiliation{Research Center for Functional Materials, National Institute for Materials Science, 1-1 Namiki, Tsukuba 305-0044, Japan}

\author{Takashi Taniguchi}
\affiliation{International Center for Materials Nanoarchitectonics, National Institute for Materials Science, 1-1 Namiki, Tsukuba 305-0044, Japan}

\author{Moty Heiblum}
\affiliation{Braun Center for Submicron Research, Department of Condensed Matter Physics, Rehovot, Israel}

\author{Cyprian Lewandowski}
\affiliation{Department of Physics, Florida State University, Tallahassee, Florida 32306, USA}
\affiliation{National High Magnetic Field Laboratory, Tallahassee, Florida 32310, USA}

\author{Mitali Banerjee}
\email{mitali.banerjee@epfl.ch}
\affiliation{Institute of Physics, École Polytechnique Fédérale de Lausanne (EPFL), CH-1015 Lausanne, Switzerland}
\affiliation{Center for Quantum Science and Engineering (QSE Center), École Polytechnique Fédérale de Lausanne (EPFL), CH-1015 Lausanne, Switzerland}

\date{\today}

\begin{abstract}
Superconductivity in rhombohedral graphene has been observed across many layer numbers, with mounting evidence pointing toward spin-triplet pairing, yet complementary probes of the superconducting spin structure remain needed. Here we use one-sided WS$_2$ proximity in rhombohedral pentalayer graphene (R5G) as a surface-selective spin-orbit probe. The induced Ising spin-orbit coupling is strongest for carriers localized near the WS$_2$ interface, allowing the displacement field to tune the overlap between superconducting carriers and the spin-orbit perturbation. We observe a strongly asymmetric superconducting landscape: two robust pockets, SC1 and SC2, survive only on mutually opposite signs of displacement field, while a third pocket, SC3, is substantially weaker. Gate-tracking features, quantum oscillations, and self-consistent band-structure calculations identify the layer polarization and Fermi-surface character of the relevant carriers. The robust superconducting states are absent or strongly weakened when the active high-DOS carriers are polarized toward the WS$_2$ interface. Since Ising spin-orbit coupling is compatible with time-reversed spin-singlet pairing but competes with same-spin intervalley triplet pairing by canting or pinning the parent spin texture, this surface-selective suppression provides additional evidence for spin-triplet superconductivity involving both hole-like and electron-like carriers. Our results establish one-sided TMD proximity as a displacement-field-tunable probe of superconducting spin structure in rhombohedral graphene.
\end{abstract}

\maketitle
\pagestyle{plain}        
\thispagestyle{plain}    

\begin{figure*}[!t]
    \centering
    \includegraphics[width=1\linewidth]{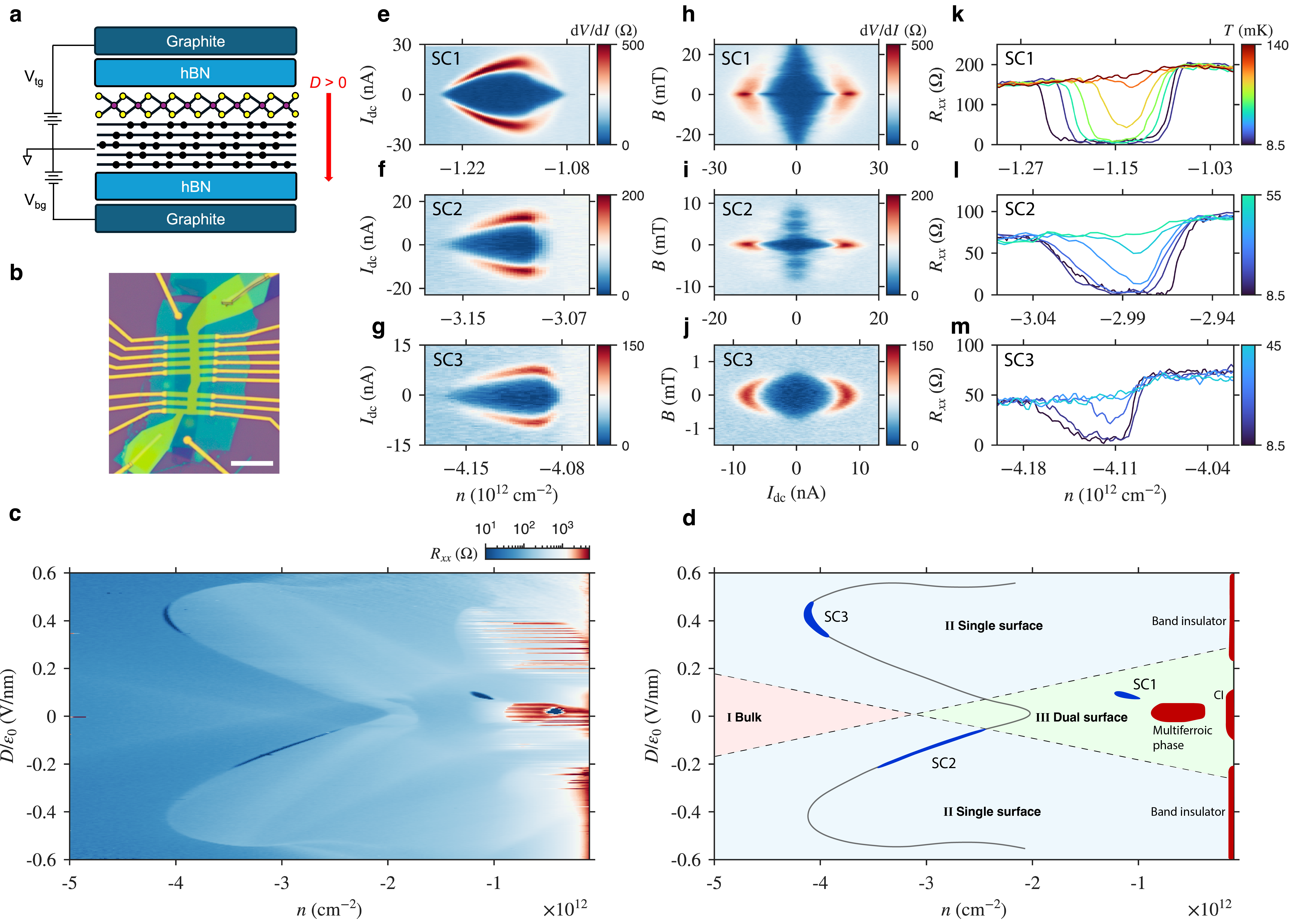} 
    \caption{\textbf{Superconductivity in WS$_2$-proximized R5G.}
    \textbf{a,} Longitudinal resistance $R_{xx}$ as a function of carrier density $n$ and displacement field $D/\varepsilon_0$ at $B=0\,\mathrm{T}$ and $T\simeq 8.5\,\mathrm{mK}$. 
\textbf{b,} Schematic of the phase diagram in \textbf{a}, labeling the bulk regime (Region~I), single-surface regime (Region~II), dual-surface regime (Region~III), the multiferroic phase, correlated insulating (CI) and band-insulating (BI) states, and the three superconducting pockets SC1--SC3. 
\textbf{c,} Device schematic: rhombohedral pentalayer graphene (R5G) proximized on one side by monolayer WS$_2$, encapsulated by hBN, and contacted by top and bottom graphite gates. The red arrow denotes the positive-$D$ convention. 
\textbf{d,} Optical micrograph of the device; scale bar, $10\,\mathrm{\mu m}$. 
\textbf{e--g,} Differential resistance $\mathrm{d}V/\mathrm{d}I$ as a function of DC current bias $I_{\mathrm{dc}}$ and carrier density $n$ for SC1, SC2, and SC3, respectively. 
\textbf{h--j,} $\mathrm{d}V/\mathrm{d}I$ as a function of perpendicular magnetic field $B$ and $I_{\mathrm{dc}}$ for SC1, SC2, and SC3, respectively. 
\textbf{k--m,} Temperature dependence of $R_{xx}$ line cuts at fixed $D$ through SC1, SC2, and SC3, respectively. 
All three pockets display nonlinear current-bias response, suppression by perpendicular magnetic field, and temperature-dependent resistance minima consistent with superconductivity.}
    \label{fig1}
\end{figure*}

\section{Introduction}

Rhombohedral $N$-layer graphene (RNG) has emerged as an exceptionally clean and gate-tunable platform for studying interaction-driven and topological electronic phases. Its low-energy surface bands can become extremely flat, producing a high density of states (DOS), while the underlying chiral band structure gives rise to substantial valley-resolved Berry curvature near the $K$ and $K'$ points of the Brillouin zone\cite{Koshino2009, Zhang2010}. Together, these ingredients promote interaction-driven symmetry breaking, magnetism, and superconductivity. In moir\'e RNG devices, alignment to a cladding layer of hexagonal boron nitride (hBN) has enabled electrically tunable integer and fractional Chern insulators at specific moir\'e fillings\cite{Lu2024, Han2024, Lu2025, Waters2025, Xie2024}. Even in crystalline, moir\'e-less devices, RNG hosts a growing family of superconducting states that are sensitive to layer number, carrier density, displacement field, and magnetic field\cite{Choi2025, Qin2026, Liu2026, Nguyen2025, Kumar2026, Guo2025}. Many of these superconductors exhibit unconventional phenomenology, including signatures of chiral pairing, large Pauli-limit violations, and superconductivity that is robust against, or even nucleated by, magnetic fields\cite{Zhou2021SC,Han2025,Seo2026,Kumar2025,Deng2026FieldSC, Xie2026FieldSC,Clogston1962,Chandrasekhar1962}. Despite this progress, the spin structure of the superconducting order parameters and its relation to the pairing mechanism remain central open questions.

A direct probe of spin structure using magnetic fields alone is challenging because several distinct microscopic mechanisms can produce field-robust superconductivity in graphene-based systems \cite{WHH1966, Maki1966,Klemm1975,FuldeFerrell1964, LarkinOvchinnikov1964, Ma2025Hc2, Kumar2025, Deng2026FieldSC, Xie2026FieldSC}. Proximity to a transition metal dichalcogenide (TMD) offers a complementary route: it induces spin-orbit coupling (SOC) in graphene, dominated by an Ising-type component that locks opposite out-of-plane spin orientations to the two valleys\cite{Zhang2023, Holleis2025, Li2024, Patterson2025, Zhang2025, Yang2025, Wang2015, Gmitra2017, Island2019,Naimer2021,GmitraFabian2015}. In a multilayer system, this proximity-induced SOC is strongest on the graphene layer adjacent to the TMD. Displacement-field control of the layer polarization therefore tunes the overlap between the superconducting carriers and the TMD interface. A singly proximitized device can thus act as a surface-selective probe of superconducting spin structure: within the same device, superconductivity can be compared when the relevant low-energy carriers are polarized either toward or away from the TMD interface, in a setting where the corresponding crystalline RNG system would otherwise be approximately symmetric under reversal of the displacement field.

Here we use this principle in rhombohedral pentalayer graphene (R5G) proximized on one side by a single layer of WS$_2$. We observe three superconducting pockets in the hole-doped phase diagram. Two robust pockets, SC1 and SC2, occur on mutually opposite sides of the displacement field compared with the positions where analogous superconductivity would be expected in pristine R5G, while a third pocket, SC3, is substantially weaker in critical temperature, critical current, and out-of-plane critical field. 

Combining magnetotransport, quantum-oscillation analysis, and self-consistent band-structure calculations, we show that the robust superconducting states are absent or strongly weakened when the relevant pairing carriers are polarized toward the WS$_2$ interface, where the induced Ising SOC is strongest. A single proximal TMD layer therefore serves as a displacement-field-tunable probe of superconducting spin structure in rhombohedral graphene, supplementing existing magnetic-field-based evidence for triplet pairing.

\section*{Superconductivity in WS$_2$-proximitized R5G}

We first characterize our rhombohedral pentalayer graphene device proximized on one side by monolayer WS$_2$. Fig.~\ref{fig1}a,b show a schematic and an optical micrograph of the device. The WS$_2$-proximitized R5G stack is encapsulated by hBN and contacted by top and bottom graphite gates, allowing independent control of the carrier density $n$ and displacement field $D$, with the WS$_2$ layer located on the top surface of the graphene stack. Fig.~\ref{fig1}c shows the longitudinal resistance $R_{xx}$ as a function of $n$ and $D/\epsilon_0$ at $B=0$ and $T\simeq 8.5\,\mathrm{mK}$. The corresponding schematic in Fig.~\ref{fig1}d labels the characteristic features of crystalline R5G, including the highly resistive correlated state near charge neutrality, the multiferroic phase near $D=0$ on the hole-doped side, and the band-insulating states at large $|D|$\cite{Han2023, Han2024, Liu2024, Han2025}. In addition, we observe three superconducting pockets, denoted SC1--SC3.

The superconducting character of all three pockets is established by nonlinear differential-resistance measurements (Fig.~\ref{fig1}e--g), their response to a perpendicular magnetic field (Fig.~\ref{fig1}h--j), and their temperature dependence (Fig.~\ref{fig1}k--m). SC1 is the most robust pocket, with a perpendicular critical field exceeding $B_{\mathrm{crit}}\simeq 26\,\mathrm{mT}$, a critical current $I_{\mathrm{dc,crit}}\simeq 24\,\mathrm{nA}$, and a transition temperature of approximately $120\,\mathrm{mK}$. SC2 is weaker, with $B_{\mathrm{crit}}\simeq 10\,\mathrm{mT}$, $I_{\mathrm{dc,crit}}\simeq 16\,\mathrm{nA}$, and superconductivity suppressed by $T\simeq 55\,\mathrm{mK}$. SC3 is the weakest of the three, with $B_{\mathrm{crit}}\simeq 0.6\,\mathrm{mT}$, $I_{\mathrm{dc,crit}}\simeq 10\,\mathrm{nA}$, and $T_c\simeq 35\,\mathrm{mK}$. The current- and field-driven nonlinearities, together with the vanishing of both $R_{xx}$ and $R_{xy}$ at zero field, distinguish these pockets from anomalous metallic states reported in other rhombohedral graphene devices\cite{Mattanometal2026}.

The key feature of the phase diagram is that superconductivity is strongly asymmetric under reversal of the displacement field. In pristine crystalline R5G, superconducting pockets appear approximately symmetrically at positive and negative $D$\cite{Han2025, Seo2026}. By contrast, in the WS$_2$-proximitized device, the robust pockets SC1 and SC2 appear only on one side of the phase diagram, while their expected counterparts at the opposite sign of $D$ are absent or strongly weakened. SC3 forms a much weaker superconducting pocket in the single-surface regime. As summarized schematically in Fig.~\ref{fig1}d, SC1 lies in the dual-surface regime, where low-energy carriers occupy states localized near both outer graphene surfaces, whereas SC2 and SC3 lie in single-surface regimes, where the relevant low-energy carriers are primarily localized near one outer surface. This displacement-field-selective hierarchy of superconducting robustness motivates the gate-tracking and quantum-oscillation analysis that follows, which identifies the vertical charge distribution and Fermi-surface character underlying each pocket.

\begin{figure*}[!t]
    \centering
    \includegraphics[width=1\linewidth]{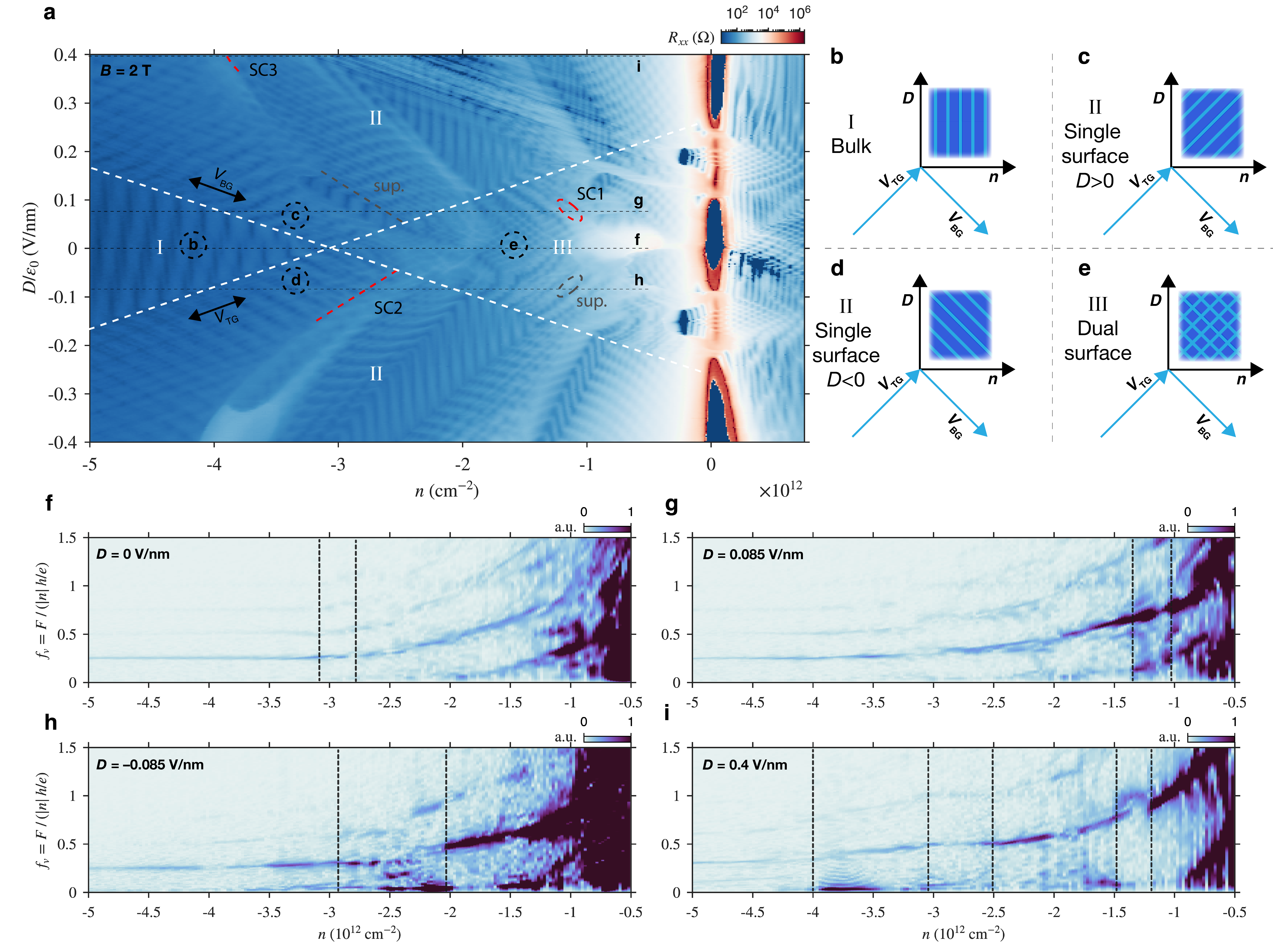} 
    \caption{\textbf{Quantum oscillations and gate-tracking features of WS$_2$-proximitized R5G.}
\textbf{a,} Longitudinal resistance $R_{xx}$ as a function of carrier density $n$ and displacement field $D/\varepsilon_0$ at $B=2\,\mathrm{T}$. On the hole-doped side, the phase diagram separates into the bulk regime (Region~I), single-surface regime (Region~II), and dual-surface regime (Region~III), bounded by gate-tracking features (white dashed lines). The superconducting pockets SC1 and SC2 (red dashed outlines), their suppressed or strongly weakened counterparts (``sup.''), and SC3 are marked. Black arrows indicate the directions of the bottom- and top-gate voltage axes, $V_{\mathrm{BG}}$ and $V_{\mathrm{TG}}$; letters denote the positions corresponding to the schematics in \textbf{b--e}. 
\textbf{b--e,} Schematic orientation of quantum-oscillation fringes in the $(V_{\mathrm{BG}},V_{\mathrm{TG}})$ plane. 
\textbf{b,} In the bulk regime, fringes are approximately orthogonal to lines of constant carrier density, indicating that the delocalized Fermi surface is tuned by both gates. 
\textbf{c,d,} In the single-surface regime at \textbf{c,} $D>0$ and \textbf{d,} $D<0$, the fringes rotate toward a single-gate-tracking orientation, indicating that the relevant Fermi-level pocket is controlled predominantly by one gate. Reversing $D$ reverses the layer polarization and transfers the surface pocket to the opposite outer graphene surface. 
\textbf{e,} In the dual-surface regime, low-energy states on both outer surfaces can be tuned differently by the two gates, producing a mixed or checkered quantum-oscillation pattern. 
\textbf{f--i,} Normalized quantum-oscillation frequency $f_v=F/(\lvert n\rvert h/e)$ versus $n$ at different fixed values of $D$. Black dashed lines serve as guides to the eye for relevant phase transitions discussed in the main text.}
    \label{fig2}
\end{figure*}

\section{Gate tracking and normal-state fermiology}

To analyze the nature of the superconducting pairing, we first need to identify the structure of the normal state, in
particular its vertical charge distribution and Fermi-surface character. On the majority hole-doped side, the phase
diagram separates into three regimes, labeled in Fig.~\ref{fig1}d and Fig.~\ref{fig2}a: a bulk regime (Region~I), a
single-surface regime (Region~II), and a dual-surface regime (Region~III). The boundaries between these regimes appear
as resistive features that track one of the two gate voltages rather than lines of constant total
density\cite{Kolar2026, Kumar2026, Guo2025,Liu2026,Yin2019,Mazzucca2025}. This single-gate tracking is a characteristic consequence of electrostatic
screening in rhombohedral multilayers. The resistive features mark the appearance of a large DOS surface-polarized pocket near $\mathbf k=0$ at the Fermi level, and are gate tracking because 
the effect of the remote gate is screened either by bulk-like states ($I \leftrightarrow II $ boundary), 
or by the surface pocket at the opposite surface ($II \leftrightarrow III $ transition). Experimentally, the vertical charge structure
can be inferred from the orientation of quantum-oscillation fringes in the $(V_{\rm BG}, V_{\rm TG})$ plane, which,
together with microscopic modelling, provides a diagnostic of whether the Fermi-level states are bulk-like, localized
near a single surface, or distributed over both outer surfaces. This is done in what follows.

In Region~I, the quantum-oscillation fringes are approximately orthogonal to lines of constant carrier density, indicating that both gates tune the same delocalized Fermi surface. The dominant normalized oscillation frequency,
\[
f_v=\frac{F}{|n|h/e},
\]
is close to \(0.25\), as expected for a fourfold-degenerate metal in which the occupied carriers are distributed among spin and valley flavors. We therefore identify Region~I as a bulk-like full metal. This experimental assignment agrees with self-consistent calculations of the band structure and real-space layer polarization, shown in Fig.~\ref{fig3}h to the left and schematically depicted in Fig.~\ref{fig3}c.

At a finite displacement field, crossing the first gate-tracking boundary brings the system into Region~II, where localized surface flat-band pockets coexist with more delocalized states. In this regime, the oscillation fringes rotate away from the constant-density direction and become approximately perpendicular to a single gate axis, as illustrated in Fig.~\ref{fig2}c,d.  This behavior again signals that the effect of a single gate is screened. However, in contrast to the gate screening responsible for the zero-field resistive features that mark the boundary between region I and II, where the feature is due to the surface states, this time the screened gate is the proximate one to the surface states, and the feature (quantum oscillations) is due to the bulk states. 
Indeed, the bulk states are much more dispersive, leading to much more robust quantum oscillations compared to those of the high DOS surface pocket. The normalized oscillation frequency of \(f_v=0.25\) indicates that the system
remains in a full metallic state close to the gate-tracking feature (see Fig.~\ref{fig2}g,h).  Reversing the sign of
\(D\) reverses the layer polarization, transferring the low-energy surface pocket from one outer graphene surface to the
other. These conclusions are supported by the self-consistent calculations in the middle of Fig.~\ref{fig3}h, which show a Fermi
surface built from single-surface carriers coexisting with delocalized bulk-like states, and the corresponding charge
distributions are schematically depicted in Fig.~\ref{fig3}d,e (depending on the sign of \textit{D}). 
Analysis of the quantum-oscillation frequencies within Region~II then distinguishes the flavor polarization and
Fermi-surface topology of these single-surface states. Depending on density and displacement field, we find full,
partially isospin-polarized (PIP), half-metallic, annular, and quarter-metallic regimes, as summarized in
Fig.~\ref{fig3}a (see Fig.~\ref{fig2}i and additional fermiology analysis in the Extended Data section). In this
normal-state map, SC2 and SC3 both lie in the single-surface regime, but on opposite sides of the displacement field.
SC2 sits at the border between a full metal and a PIP phase (see Fig.~\ref{fig2}h), while SC3 sits at the border between
an annular full metal and a PIP phase (see Fig.~\ref{fig2}i). We note that oscillations in the flavor-polarized half- and quarter-metallic regimes can again appear approximately orthogonal to lines of constant density. This naturally arises when flavor ferromagnetism moves the surface polarized high DOS states away from the Fermi level, disabling screening, and leaving only states with a single vertical distribution which respond to the total density only. We therefore use the normalized frequencies mainly to infer flavor degeneracy and Fermi-surface topology, while assigning surface polarization from the surrounding gate-tracking features and self-consistent band-structure calculations.

At smaller \(|D|\) and lower \(|n|\), the system crosses a second gate-tracking boundary into Region~III, where both outer surfaces host low-energy flat-band states and contribute to screening. The quantum-oscillation pattern is therefore more complex because the two surfaces can be tuned differently by the two gates (see checkered quantum oscillations pattern at a representative point in Region~III in Fig.~\ref{fig2}e) and can undergo independent or coupled isospin symmetry breaking. At \(D=0\), the two outer surfaces are related by the approximate mirror symmetry of the crystalline stack, so the charge distribution has no net top-bottom layer polarization, but for any finite \(|D|\), however, carriers in this regime occupy low-energy states localized near opposite outer graphene surfaces, as shown schematically in Fig.~\ref{fig3}f,g. The associated quantum-oscillation spectrum is consistent with an approximately unpolarized, full-metal degeneracy (See Fig.~\ref{fig2}f) and is captured by the self-consistent calculation in Fig.~\ref{fig3}i,j.

Upon further decreasing \(|n|\), interaction-driven symmetry breaking produces dual-surface states compatible with the observed oscillation spectra and real-space charge distributions shown at the positions labeled 4 and 5 in Fig.~\ref{fig3}f,g. Quantum oscillation frequencies indicate that SC1 (located at position 5) resides in an annular half-metallic phase (see Fig.~\ref{fig2}g). We consider two closely related candidate parent states for SC1 proposed previously in Refs.\cite{Kumar2026, Guo2025,Liu2026}. In one scenario, the parent state is a layer-antiferromagnetic metal, with hole-like pockets on opposite surfaces occupying different valley flavors and with spin degeneracy lifted, so that the net anomalous Hall response cancels; a representative calculation is shown in Fig.~\ref{fig3}j. In the second scenario, the parent state is an electron-hole half-metallic semimetal, in which an electron-like pocket is surface-polarized while a hole-like pocket remains more delocalized, and the remaining flavor degeneracy is reduced consistently with the observed half-metallic quantum-oscillation frequencies; a representative calculation is shown in Fig.~\ref{fig3}h to the right and in Fig.~\ref{fig3}i. Both possibilities are consistent with the absence of a large anomalous Hall signal and with the dual-surface fermiology inferred from the gate-tracking data (see Fig.~\ref{fig2}e). However, the doping dependence of the Shubnikov--de Haas oscillations favors an electron-hole semimetal interpretation with reduced spin/flavor degeneracy for the parent state of SC1, consistent with earlier experimental work\cite{Liu2026}. In this interpretation, as argued previously in Ref. \cite{Kumar2026}, an electron-like pocket is surface-polarized while the hole-like pocket remains more delocalized; the high-DOS electron-like pocket is therefore the most likely active superconducting band, while the hole-like carriers may primarily contribute through screening or interband coupling. The above normal-state assignment provides the basis for the spin-orbit-coupling analysis below.

\section{Suppression of spin-triplet superconductivity by Ising-type spin-orbit coupling}

\begin{figure*}[!t]
    \centering
        \includegraphics[width=1\linewidth]{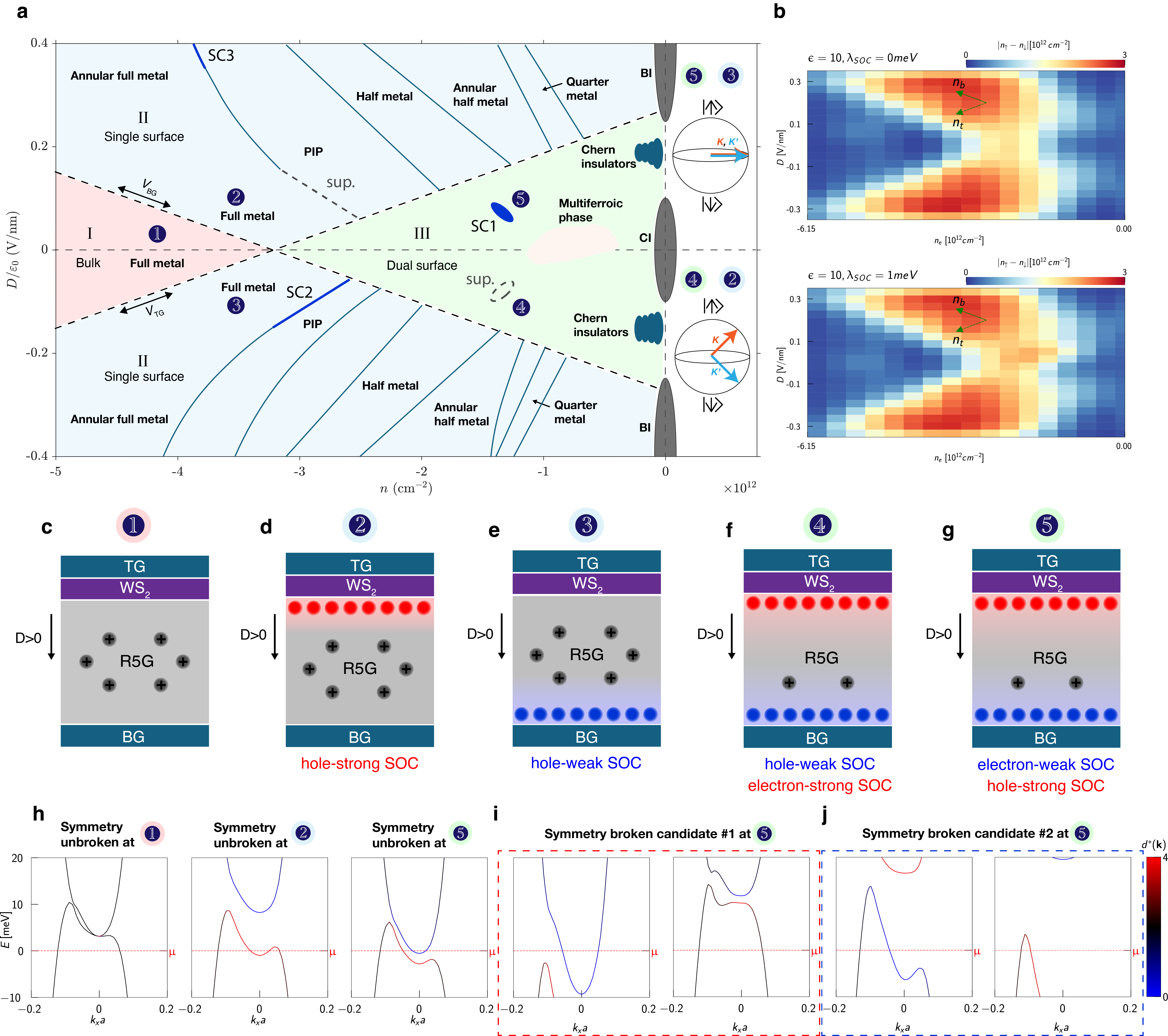} 
    \caption{\textbf{Normal-state fermiology and surface-selective spin-orbit coupling.} \textbf{a,} Schematic $n$--$D$ phase diagram of the hole-doped side obtained from the analysis of Fig.~\ref{fig2}. Region~I is a bulk-like full metal. Region~II contains single-surface metallic states, including full-metal, partially isospin-polarized (PIP), half-metallic, annular half-metallic, and quarter-metallic regimes. Region~III contains dual-surface metallic, correlated insulating (CI), and multiferroic phases, terminating in band-insulating (BI) states at large $|D|$. Circled numerals~1--5 mark representative positions shown in \textbf{c--g} and \textbf{h--j}; SC1, SC2, and SC3 are indicated. Insets illustrate the valley-contrasting Ising SOC induced by WS$_2$ for states with appreciable weight on the TMD-proximate surface.
\textbf{b,} Numerical Hartree-Fock map of flavor polarization in the $n-D$ plane illustrating the effect of increasing proximity-induced Ising SOC on the normal-state electronic structure.
\textbf{c--g,} Real-space schematics of the carrier distribution across the R5G/WS$_2$ stack for the representative positions marked in \textbf{a}.
\textbf{h,} Self-consistent electrostatic band structures without symmetry breaking. Layer polarizations are shown in color.
\textbf{i,j} Hartree-Fock band structures for a two-flavor system for the candidate
ferromagnet (i) and layer antiferromagnet (j) state.
}
\label{fig3}
\end{figure*}

We now use the normal-state assignments above to interpret the displacement-field asymmetry of the superconducting pockets. The proximity-induced SOC from WS$_2$ is localized primarily on the graphene surface adjacent to the TMD. For the realistic SOC strengths relevant here, at most a few meV, the self-consistent calculations in Fig.~\ref{fig3}b indicate that the SOC perturbation does not qualitatively reorganize the Coulomb-dominated fermiology or layer polarization. Its main role is instead to modify the spin structure of states with appreciable weight on the WS$_2$-proximate surface. This distinction allows us to treat the TMD as a surface-selective spin perturbation acting on the normal states identified in Figs.~\ref{fig2} and \ref{fig3}.

The spin-selective effect of Ising SOC should be distinguished from the band-structure reconstruction discussed above. The self-consistent band-structure calculations identify the Fermi-surface pockets, their carrier character, and their layer polarization, but they do not by themselves determine the spin texture of the symmetry-broken parent state. If the spin texture is held fixed, valley-contrasting Ising SOC shifts same-spin states in opposite valleys in opposite directions, which is incompatible with same-spin intervalley pairing (e.g. see Refs.\cite{Frigeri2004, Ma2025Hc2}). Once spin-sector interactions, such as Hund's coupling, are included, the parent state can respond differently: the two valleys may retain a common in-plane spin component while acquiring opposite out-of-plane components. The valley quasiparticle bands can then remain degenerate, but the common same-spin component relevant for intervalley triplet pairing is reduced as the state is pushed toward spin-valley locking. The extent to which the spin texture is controlled by Ising SOC is therefore set by the competition between the projected Ising SOC and the Hund/canting stiffness of the parent metal. Thus, the surface-selective suppression observed here should be viewed as evidence that Ising SOC competes with the same-spin triplet instability, either by reducing the compatible spin component of the parent state or by stiffening the associated spin-canting modes, rather than necessarily by introducing a literal noninteracting valley energy mismatch.

The above distinction is also consistent with the in-plane-field-induced superconductivity reported in rhombohedral graphene without a proximitizing TMD\cite{Kumar2025}. There, an in-plane magnetic field acts primarily as a uniform Zeeman perturbation and can favor an in-plane spin-polarized parent state compatible with same-spin triplet pairing. The WS$_2$-induced Ising SOC considered here is different: it is a valley-contrasting out-of-plane spin-orbit field localized on one surface. Rather than acting as a uniform spin polarizer that enhances a common in-plane spin alignment, as in the field-induced or field-stabilized superconducting regimes of pristine rhombohedral graphene\cite{Kumar2025}, the Ising SOC competes with the common in-plane spin orientation of the two valleys. This provides a natural way to understand why the field-induced superconducting ``river'' reported in pristine rhombohedral graphene may be absent or strongly modified in the present WS$_2$-proximitized device.

We first consider the single-surface superconductors in Region~II. SC2 occurs where the quantum-oscillation and gate-tracking analysis identifies a single-surface hole-like pocket localized away from the WS$_2$ interface, as depicted in Fig.~\ref{fig3}e. The corresponding positive-$D$ configuration places the analogous hole-like pocket near the WS$_2$ interface, Fig.~\ref{fig3}d, and the robust superconducting state is absent. This displacement-field selectivity is consistent with a hole-like same-spin triplet superconductor whose parent spin texture is destabilized when the active pocket is strongly exposed to Ising SOC. SC3 lies nearby in the single-surface regime but is much weaker, with substantially reduced $T_c$, critical current, and perpendicular critical field. We therefore do not interpret SC3 as restoring the approximately symmetric superconducting phase diagram of pristine R5G. Rather, SC3 shows that superconductivity is not necessarily eliminated completely in the SOC-exposed single-surface regime, but the robust instability associated with SC2 is strongly suppressed.

We next consider SC1, which emerges from the dual-surface Region~III. As discussed above, the parent state of SC1 is most naturally interpreted as an electron-hole semimetal with reduced spin/flavor degeneracy\cite{Kumar2026, Liu2026, Guo2025}, although a layer-antiferromagnetic metallic state remains a nearby alternative normal-state configuration. In both cases, the absence of a large anomalous Hall response disfavors a purely valley-polarized parent and instead points to a state in which the relevant low-energy carriers occupy both valleys. When combined with the half-metallic quantum-oscillation frequencies (see Fig.~\ref{fig2}g,h), this suggests a parent with spin polarization and two-valley participation, the normal-state setting expected to favor equal-spin intervalley pairing. In the electron-hole semimetal interpretation, the calculated band structure in Fig.~\ref{fig3}h to the right and in Fig.~\ref{fig3}i indicates the presence of a high-DOS electron-like pocket and a more delocalized hole-like pocket. Experimentally, SC1 survives when the electron-like pocket is localized away from WS$_2$, and is absent or strongly weakened when the corresponding pocket is moved toward the TMD. Thus, although SC1 emerges from a dual-surface normal-state regime, the superconducting instability appears to be dominated by the surface-polarized high-DOS electron-like pocket, with the hole-like carriers likely contributing primarily through screening or interband coupling\cite{Kumar2026, Guo2025}. The same surface-selective suppression pattern therefore points to same-spin triplet pairing also for the electron-like superconducting state associated with SC1.

Taken together, SC1 and SC2 show complementary realizations of the same diagnostic: robust superconductivity is observed when the active high-DOS carriers are weakly exposed to WS$_2$-induced Ising SOC, while the corresponding state is absent or strongly weakened when those carriers acquire strong weight on the TMD-proximate surface. SC3 fits within this hierarchy as a weak residual superconducting pocket in the single-surface regime. This phenomenology is difficult to reconcile with a conventional spin-singlet interpretation, for which Ising SOC is not expected to be pair-breaking in the same way, and instead provides strong evidence for same-spin triplet superconductivity involving both hole-like and electron-like carriers in rhombohedral graphene.

\section{Concluding remarks}

Our results establish one-sided TMD proximity in RNG as a surface-selective probe of superconducting spin structure. The central observation is not that WS$_2$ suppresses superconductivity uniformly across the $n$--$D$ phase diagram, but that robust superconductivity is absent or strongly weakened when the relevant high-DOS carriers are polarized toward the WS$_2$ interface. This selectivity helps identify the Fermi-surface pockets most closely associated with pairing, and complements existing evidence for spin-triplet superconductivity in RNG, which has primarily relied on magnetic-field response, Pauli-limit violation, and field-induced or field-stabilized superconductivity.

SC1 in particular highlights the richness of the dual-surface regime and the insights gained from our analysis. Although SC1 emerges from Region~III, the superconducting instability appears to be dominated by the surface-polarized, high-DOS electron-like pocket, while the accompanying hole-like carriers are more delocalized\cite{Kumar2026, Liu2026, Guo2025}. Thus, SC1 is best viewed as superconductivity emerging from a dual-surface normal-state environment, not necessarily as a condensate with equal weight on both surfaces. We note that possible band inversions, as suggested in Refs.~\cite{Han2024QAH, Liu2026}, further caution against identifying the active surface from the electron- or hole-like character alone. For our spin-orbit interpretation, the relevant quantity is the self-consistent layer polarization of the active pocket, rather than its strict carrier type.

The locations of the superconducting pockets suggest that pairing is tied to nearby normal-state reconstructions, not simply to the presence of a high DOS. SC2 appears within the single-surface regime, in a region where nearby quantum-oscillation features indicate changes in Fermi-surface topology and flavor polarization, while SC1 emerges near the dual-surface electron-hole regime. The termination of SC2 near Region~III and the absence of a robust mirrored pocket on the TMD-proximate side may therefore reflect more than pair breaking of an otherwise unchanged parent state. In the present device, the one-sided Ising SOC also acts as an additional symmetry-breaking perturbation that can modify the competition among nearby spin- and valley-polarized metallic states. SC3 likely belongs to the same hierarchy: it is much weaker, displaced toward larger $|D|$, and may reflect a local fermiology in which the band structure, screening environment, or interaction-renormalized surface polarization has shifted relative to pristine R5G.

Our interpretation is also consistent with earlier experiments in which TMD-induced SOC enhances or nucleates superconductivity\cite{delaBarrera2018Ising,Lu2015IsingMoS2}. In Bernal bilayer graphene proximitized by WSe$_2$ (e.g. Refs.~\cite{Zhang2023, Holleis2025, Li2024, Chou2022BLGWSe2}), Ising SOC appears to promote superconductivity when the relevant carriers are pushed toward the TMD interface, consistent with enhancement of pairing between time-reversed valley partners, such as opposite-spin intervalley pairs. Rhombohedral systems show a complementary phenomenology. In pristine rhombohedral graphene, in-plane magnetic field can induce or stabilize superconductivity, consistent, though not by itself definitive, with a spin-polarized parent state favorable to same-spin triplet pairing with a common in-plane spin component\cite{Kumar2025}. Recent work on bare R4G and R5G further shows that magnetic fields and proximitized SOC can enhance, suppress, or generate different superconducting pockets depending on the underlying spin/valley parent state and pairing channel\cite{Seo2026}. In spin-orbit-proximitized rhombohedral trilayer graphene, superconductivity appears near a transition between a spin-canted state with finite in-plane moment and a fully spin-valley-locked state, reflecting competition between Hund's coupling and Kane--Mele/Ising SOC (e.g. Refs.~\cite{Patterson2025, Dong2026,KohAlicea2024}). Our result probes a different limit of the same spin physics: one-sided WS$_2$ proximity introduces a valley-contrasting out-of-plane field only for carriers with weight on the TMD-proximate surface. This field need not create a literal valley energy mismatch in an interacting spin-polarized parent. Instead, it can reduce the common in-plane spin component compatible with same-spin intervalley pairing and thereby suppress the triplet instability.

Finally, the observed hierarchy of superconducting robustness may reflect the competition between projected Ising SOC and the effective Hund/canting stiffness of the parent metal. The bare proximity SOC is set by the WS$_2$ interface, but the effective SOC seen by an active band depends on its TMD-surface weight, which is determined by the fermiology and layer polarization discussed above. By contrast, the Hund/canting scale is an emergent spin-sector energy that can vary with experimental tuning parameters such as carrier density and displacement field. Weak pockets such as SC3 may therefore mark an intermediate regime in which the projected Ising SOC is strong enough to weaken the robust triplet instability, but local exchange/canting physics and fermiology still allow a fragile superconducting state to survive. In this canted regime, Ising SOC does not simply produce a valley energy mismatch; it pins the spin sector and can stiffen or gap spin-canting modes that may otherwise remain soft and contribute to pairing. This interpretation is closely related to the proposed mechanism for the $B_\perp$ enhancement of SC3 in bare R5G, where weak intrinsic Kane--Mele SOC and Hund-dominated in-plane spin polarization allow a small out-of-plane field to generate spin canting, valley imbalance, and a DOS enhancement for an intravalley pairing channel\cite{Seo2026}. In the present one-sided WS$_2$ device, the much larger projected Ising SOC on the TMD-proximate surface places the system in a different limit of the same canting energetics, where spin pinning can suppress a same-spin intervalley triplet instability rather than enhance an intravalley one. A related possibility is suggested by recent work on rhombohedral graphene on WSe$_2$, where neighboring zero-resistance superconducting and anomalous-metal pockets display similar onset phenomenology but distinct low-temperature resistance, critical-field scales, and current response\cite{Mattanometal2026}. Future measurements and theoretical analysis of the in-plane critical field, superfluid stiffness, pairing gap, and spin-valley collective modes should help distinguish between these possibilities.

\textbf{Acknowledgments}\\
We are grateful to Matt Yankowitz and \'Etienne Lantagne-Hurtubise for helpful discussions and a critical reading of the manuscript. We also acknowledge helpful discussions with Titus Neupert. K.K. acknowledges Emily Hajigeorgiou and Nanyu Yao for their help in upgrading the custom cold finger with integrated low-temperature filtering at the sample stage, substantially reducing the electron temperature of the measurement setup. K.K. and Z.Z. acknowledge funding from SNSF. M.B. acknowledges the support of the SNSF Eccellenza grant No. PCEGP2\_194528. K.W. and T.T. acknowledge support from the JSPS KAKENHI (Grant Numbers 20H00354 and 23H02052) and World Premier International Research Center Initiative (WPI), MEXT, Japan. C. L. was supported in part by start-up funds from Florida State University and the National High Magnetic Field Laboratory and in part by NSF CAREER grant No. DMR-2543710. The National High Magnetic Field Laboratory is supported by the National Science Foundation through NSF/DMR-2128556 and the State of Florida. K. Ko. was supported by a grant from the Simons Foundation (SFI-MPS-NFS-00006741-12, P.T.) in the Simons Collaboration on New Frontiers in Superconductivity. M.H. acknowledges the support of the Israel Science Foundation, Grant No. 1510/22.

\textbf{Author contributions}\\
M.B., K.K., and Z.Z. conceived the project. M.B. supervised the project. Z.Z. and Y.L. fabricated the device. K.K., Z.Z., and Y.L. performed the measurements at EPFL. Z.Z. performed the measurements at WIS. K.K. and Z.Z. analyzed the data with input from M.B., K.Ko., and C.L. K.Ko. performed the theoretical calculations with input from C.L. M.H. provided access to the dilution refrigerator at WIS. K.W. and T.T. provided the hBN crystals. K.K., C.L., and M.B. wrote the manuscript with inputs from all authors.

\textbf{Data availability}\\
The data supporting the findings of this study are available from the corresponding author upon reasonable request.

\textbf{Methods}\\
\textbf{Device fabrication}\\
Graphite/graphene and hBN were first exfoliated onto an O$_2$-plasma cleaned SiO$_2$-coated (285 nm) Si substrate, while the WS$_2$ was exfoliated onto PDMS, followed by detailed optical microscope characterization. Suitable homogeneous hBN flakes for encapsulation ($\approx$ 30 nm) and graphite-flakes for gating ($\approx$ 2–3 nm) were selected and their quality confirmed via AFM imaging. The layer number of the pentalayer graphene and the WS$_2$ flakes was determined via calibrated optical contrast, and the rhombohedral (RH) domain was resolved via infrared microscopy. The selected RH domain was isolated via anodic oxidation in an AFM using a conductive tip to reduce the chance of relaxation into a Bernal stacking order during fabrication. An auxiliary hBN layer, the bottom hBN layer, and the bottom graphite gate were picked up employing an all dry-transfer method by using a poly(bisphenol A-carbonate) (PC) covered PDMS stamp with a 2D material transfer station. The auxiliary hBN layer was removed by dragging it off the bottom hBN layer with a specialized small contact point PDMS stamp. The top graphite gate, the top hBN layer, the WS$_2$, and the pentalayer graphene were picked up with a poly(bisphenol A-carbonate) (PC) covered PDMS stamp and released onto the prepared bottom part. The device structure was created with standard electron beam lithography and reactive ion etching techniques. The electrical connections were established with e-beam evaporation of Cr/Au (5~nm/70~nm)
\\

\textbf{Electrical transport measurements}\\
The measurements were carried out in two separate dilution refrigerators at EPFL and WIS with base phonon temperatures of 8–12 mK. The longitudinal and transverse resistances were measured with SRS and Zurich Instruments MFLI lock-in amplifiers at an excitation current of 0.5–5 nA, demodulated at a frequency of either 17.777 Hz or 73.777 Hz. A cold ground shorted to the cold finger at the sample stage was used to further reduce the electronic temperature. The gate voltages were applied using Yokogawa GS200 and Keithley 2400 source-meters. The carrier density and displacement field were determined according to $n=(C_\text{TG}V_\text{TG}+C_\text{BG}V_\text{BG})/e$ and $D/\varepsilon_0=(C_\text{TG}V_\text{TG}-C_\text{BG}V_\text{BG})/2\varepsilon_0$, respectively, where $C_\text{TG/BG}$ and $V_\text{TG/BG}$ are the top and bottom gate capacitances per unit area and voltages, respectively. The gate capacitances depend on the hBN layer thicknesses and were determined via Hall resistance measurements at different gate-defined carrier densities. 

\textbf{Details of theoretical analysis}

\textit{Rhombohedral electrostatic model }
\def\dens{{n}}
\newcommand{\proj}{\mathcal P_{\text{active}} }
\def\nactive{N_{\text{active}}}
\def\nflav{N_{\text{flavors}}}
\def\ndeg{N_{\text{deg.}}}

We model rhombohedral graphene by considering the following single-particle Hamiltonian
\cite{mccann_landau-level_2006,nilsson_electronic_2008,mccann_electronic_2013}

\begin{equation}
\hat H_{SP}=
\label{eq:hambernal}
\sum_{\mathbf k}
\Psi^\dagger_{\mathbf k}
    \begin{pmatrix}
        U_1 &  v_F \overline{k}&-v_4 \overline{k}&-v_3 k &0&t_2& \\
   v_F k& U_1&t_1  &-v_4 \overline{k} &\\ 
         -v_4 k& t_1&U_2 &  v_F \overline{k}& \\
        -v_3  \overline{k}&-v_4 k& v_F k &U_2&  \\
             0& &&& \ddots\\

        t_2& &&&& \ddots\\
        \vdots&  
    \end{pmatrix}
\Psi_{\mathbf k} 
\end{equation}
the basis $\Psi_{\mathbf k} = (c_{\mathbf k,1,A}, c_{\mathbf k,1,B}, c_{\mathbf k,2,A}, c_{\mathbf k,2,B}, \dots)$, 
where $k=k_x+ik_y$, $\overline{k}=k_x - ik_y$ momenta are measured from the K point, $v_F$ is the graphene Dirac velocity,
and $v_3, v_4 \ll v_F$ denote nonlocal interlayer tunneling velocities, and $t_1$ is the strength of $B1\to A2$ tunneling.
$U_l$ is the layer-dependent electrostatic potential.
We use the parameters following~\cite{jungParkTopologicalFlatBands2023}. Specifically, we take
$v_F = \SI{-547}{meV\cdot nm}$, $v_3=\SI{61.66}{meV\cdot nm}$ and
$v_4  = \SI{30.3}{meV\cdot nm}$, $t_1   =  \SI{356.1}{meV}$
$t_2  =  \SI{-4.15}{meV}$.

Note that we do not include additional potentials on the outer layers as we determine the layer potential fully self-consistently by electrostatics as detailed below.

This potential is determined using Gauss' law as
\begin{equation}
\label{eq:potdifference}
    U_{l+1}-U_{l} = -e^2 d \frac{\dens_b+ \sum_{j \leq l}\dens_j }{\epsilon_\perp\epsilon_0}.
\end{equation}
with $d$ the interlayer distance and $\epsilon_\perp$ the out-of-plane dielectric constant. 
We denote the net electron densities in each layer
$l$ by $\dens_l$, and in the top and bottom gates by $\dens_t$ and $\dens_b$,
respectively. 

Overall charge neutrality implies that
the sum of these gate charges fixes the total device density, $\dens = \sum_l
\dens_l = -(\dens_t + \dens_b)$. Their difference sets the experimentally
accessible displacement field, $D= e\frac{\dens_b-\dens_t}{2\epsilon_0}$

At a given combination of $\dens_t$ and $\dens_b$, we need to solve Eq.~\eqref{eq:hambernal}
together with Eq.~\eqref{eq:potdifference} self-consistently.
The solution is a density matrix
$[P(\mathbf k)]_{l,s, l',s'} = \langle 
c^\dagger_{\mathbf k,l,s }c^{\phantom{\dagger}}_{\mathbf k,l',s'}
\rangle$,
which determines the layer densities as 
$\rho_l =\frac{N_{\text{deg.}}}{A}\sum_{\mathbf k,s}  \left\{[P(\mathbf k)]_{l,s, l,s}- \frac{1}{2} \right\},$
where the factor of $1/2$ subtracts the density at charge neutrality.
Since the above Hamiltonian corresponds to the K-valley only,
we model a symmetric state with fourfold degeneracy by multiplying the density of the K-valley state by $N_{\text{deg.}}=4$.

\textit{Symmetry breaking model}
To study interaction effects beyond layer potentials, and states without perfect flavor degeneracy, 
we consider $N_{flavor}$ copies of the above single-particle Hamiltonian, modelling $N_{flavor}$ flavors as
\begin{equation}
\hat H = \sum^{N_{flavor}}_f \hat H_{SP,f} +H_{\text{int}},
\end{equation}
adding the following interaction term
\begin{equation}
H_{\text{int}} = \frac{1}{2A} \sum_{\mathbf{q}\neq 0}\sum_{l,l'} V_{ll'}(\mathbf q) :\rho_{\mathbf q,l} \rho_{-\mathbf q,l'}:,
\label{eq:interactinghamiltonian},
\end{equation}
where $V_{ll'}(\mathbf q)$ is the double-gate screened Coulomb interaction \cite{lewandowskiKolarElectrostaticFate$N$layer2023},
$A$ is the system area, $::$ denotes normal ordering, and $\hat \rho_{-\mathbf q,l}$ is the charge density operator in layer $j$ at momentum $\mathbf q$.
The double-gate screened Coulomb interaction reads \cite{lewandowskiKolarElectrostaticFate$N$layer2023}

\begin{equation}
\label{eq:doublegatecoulomb}
\begin{split}
V_{ll'}(\mathbf q)
={}& \frac{1}{2\epsilon\epsilon_0 q}
\Bigg[
e^{-q|z_l-z_{l'}|} \\
&+\frac{e^{-q(z_l+z_{l'})}}
       {e^{4d_s q}-1}
\Big(
-e^{2q(d_s+z_l+z_{l'})}
-e^{2d_s q} \\
&\hspace{4.2cm}
+e^{2qz_l}
+e^{2qz_{l'}}
\Big)
\Bigg],
\end{split}
\end{equation}

where $\epsilon$ is the dielectric constant, which we allow to differ from $\epsilon_\perp$, 
and where $d_s$ is the gate-sample distance.

Within our approach, we first determine a self-consistent
$P(\mathbf k)$ from the symmetric calculation which includes only the layer potentials.
We project on $N_{\text{active}}$ bands of this calculation closest to the Fermi level,
with wavefunctions $\ket{u_{\mathbf k, \alpha}}$, $\alpha=1,\ldots, N_{\text{active}}$, where the operator
for each band for flavor $f$ is $d^\dagger_{\mathbf k,f,\alpha }$.
Transforming from the active to orbital basis is done using the matrix
$U_{ls, \alpha}(\mathbf k) = \braket{ls| u_{\mathbf k, \alpha}}$, where $l$ is a layer index and $s$ a sublattice index,
appropriate for rhombohedral $N$-layer graphene.
The projector operator on the active bands is
$\proj(\mathbf k) = \sum_{\alpha=1}^{\nactive}\ketbra{u_{\mathbf k, \alpha}}$,
 which can be expressed in orbital basis as
 \begin{equation}
\proj(\mathbf k)|_{ls,l's'} = 
\sum_{\alpha=1}^{\nactive} U_{ls, \alpha}(\mathbf k)
U^\dagger_{\alpha,l's'}(\mathbf k).
 \end{equation}

We assume that all the bands below the $\nactive$ bands are fully filled and denote by $P^0(\mathbf k)$ the density matrix (in layer and sublattice basis) of the filled remote bands.
It can be obtained as  
\begin{equation}
P^0(\mathbf k) =   P(\mathbf k) - \proj(\mathbf k) P(\mathbf k) \proj(\mathbf k).
\end{equation}
The remote filled bands contribute an additional Fock term which we need to include and which we give below.
Written in terms of the density matrix
$[P_f(\mathbf k)]_{\alpha \beta} = \langle 
d^\dagger_{\mathbf k,f,\alpha }d^{\phantom{\dagger}}_{\mathbf k,f,\beta}  
\rangle$,
the Fock term reads
\begin{align}
 \label{eq:hmffock}
\hat H_{Fock} &=-\frac{1}{A}\sum_{f}\sum_{l,l'}\sum_{\mathbf q,\mathbf k}  
V_{l,l'}(\mathbf q) 
\nonumber \\ 
&\times d^\dagger_{\mathbf k,f}\left[
\Lambda^{l}_{\mathbf q}(\mathbf k) P^T_f(\mathbf k + \mathbf q) \Lambda^{l'}_{- \mathbf q}(\mathbf k +\mathbf q)
\right]d_{\mathbf k,f},
\end{align}
where $\Lambda^{l}_{\mathbf q}$ is the overlap matrix:

\begin{equation}
[\Lambda^{l}_{\mathbf q}(\mathbf k)]_{\alpha,\beta} =\braket{u_{\mathbf k, \alpha}|u_{\mathbf k+\mathbf q, \beta}}.
\end{equation}

We also need to include the Fock term due to the filled valence bands, whilst subtracting a background. 
Explicitly, it reads
\begin{multline}
 \label{eq:hmffockb}
\hat H^{Remote}_{Fock} =-\frac{1}{A}\sum_{f}\sum_{l,l'}\sum_{\mathbf q,\mathbf k}  
V_{l,l'}(\mathbf q) 
\nonumber \\ 
\times d^\dagger_{\mathbf k,f} U^\dagger(\mathbf k)
\left\{(P^0)^T(\mathbf k + \mathbf q) - \frac{1}{2}\delta_{l,l'}\delta_{s,s'}
\right\}U(\mathbf k) d_{\mathbf k,f}.
\end{multline}

The last component is the layer densities, which are in this projected approach given as
\begin{multline}
\rho_l =\frac{1}{A }\sum_{\mathbf k}  \Bigg\{
\frac{\ndeg}{\nflav} \sum_f \mathrm{Tr}\Big[ \Lambda^{l}_{\mathbf q=0}(\mathbf k) P^T_f(\mathbf k ) \Big]+
\\
+\ndeg \sum_{s}\left( [P^0(\mathbf k)]_{ls, ls}- \frac{1}{2}\right) \Bigg\},
    \end{multline}
    where the first term is the density of the active bands, while the second term is the remote charge
density due to the filled valence bands with the density matrix at infinite temperature subtracted.

To summarize, the mean-field Hartree-Fock Hamiltonian reads

\begin{equation}
\hat H_{\text{MF}} = \sum^{N_{flavor}}_f \hat H_{SP,f} + \hat H^{Remote}_{Fock} + \hat H_{Fock}.
\end{equation}

\textit{Effect of proximitized TMD}
To evaluate the effect of proximitized TMD,
we add a projected Ising SOC term 
\begin{equation}
\hat H_{TMD,f}  = \frac{\lambda_I}{2} (-1)^f \,\, d^\dagger_{\mathbf k,f} d_{\mathbf k,f}
\end{equation}
where we take the value of
$\lambda_I = \SI{1}{meV}$ following Ref.~\cite{Zhang2023}.

\textit{Detailed numerical parameters}
We use the optimal damping algorithm \cite{lebrisCancesCanWeOutperform2000} to aid convergence. 
For the pure electrostatic Hartree band structure calculations, we assume a $C_{3z}$ symmetric solution.
This means that we use a reduced $\mathbf k$ grid, in which points related by $C_{3z}$ symmetry are identified.
We use
a grid of 519 $\mathbf k$-points centered around $\mathbf k=0$ with radius
$|\mathbf k_{max}| = \SI{0.93}{nm^{-1}}$ 
We use an out-of-plane dielectric constant of $\epsilon_\perp=3$. 
This value is appropriate since we treat the screening of layer potentials by remote bands consistently. 

For the Fock candidate state calculations at the SC2 point,
we simulate $\nflav=2$ flavors, thus
assuming a twofold $\ndeg=2$ degeneracy. 
We do not assume $C_{3z}$ symmetry, and our $\mathbf k$ grid has 1555 $\mathbf k$ points
with radius $|\mathbf k_{max}| = \SI{0.93}{nm^{-1}}$.
Note that we do not include SOC for these candidate state calculations.

For the calculations of the effect of SOC, which
are sweeps in the $n-D$ plane, 
we use a smaller grid of just 379 $\mathbf k$ points with radius $|\mathbf k_{max}| = \SI{0.91}{nm^{-1}}$.

We use a larger in-plane dielectric constant $\epsilon=10$ 
for all Fock calculations.
This is to account for remote band screening. 
At each point in the $n-D$ plane, we find the optimal state by proposing two candidate states -- the layer antiferromagnet
and the ferromagnet. We let the iterations converge for each initial state and obtain the final phase diagram by choosing the 
lower energy solution.

\bibliographystyle{apsrev4-2}
\bibliography{references.bib}

\appendix
\setcounter{figure}{0} 
\renewcommand{\figurename}{Extended Data Figure}
\section{Extended Data}
\begin{figure*}[h]
    \centering
    \includegraphics[width=.8\linewidth]{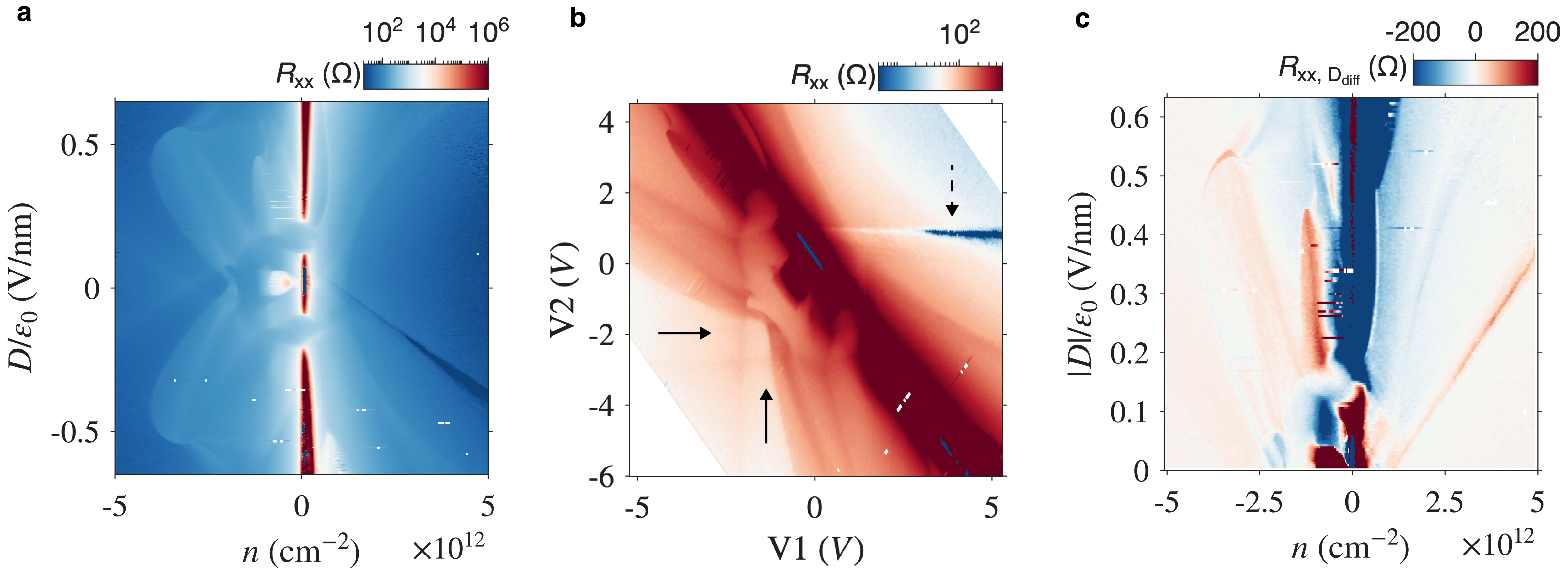} 
    \caption{\textbf{a,} Longitudinal resistance $R_{xx}$ as a function of carrier density $n$ and displacement field $D/\varepsilon_0$ at $B = 0\,\mathrm{T}$ and $T \approx 150\,\mathrm{mK}$. \textbf{b,} Longitudinal resistance $R_{xx}$ as a function of the bottom and top graphite gate voltages V1 and V2, respectively at $B = 0\,\mathrm{T}$ and $T \approx 150\,\mathrm{mK}$. The color scale is adjusted to render the gate-tracking features more visible. They are marked with black arrows. The dashed black arrow denotes a measurement artifact running parallel to the bottom gate. \textbf{c,} Difference in longitudinal resistance $R_{xx, \mathrm{D_{diff}}}$ as a function of carrier density $n$ and absolute value of the displacement field $\lvert D\rvert/\varepsilon_0$ at $B = 0\,\mathrm{T}$ and $T \approx 150\,\mathrm{mK}$. A positive $R_{xx, \mathrm{D_{diff}}}$ indicates a feature being absent on the bottom half of the $n$--$D$ map and vice versa.}
    \label{extfig1}
\end{figure*}

\begin{figure*}[h]
    \centering
    \includegraphics[width=1\linewidth]{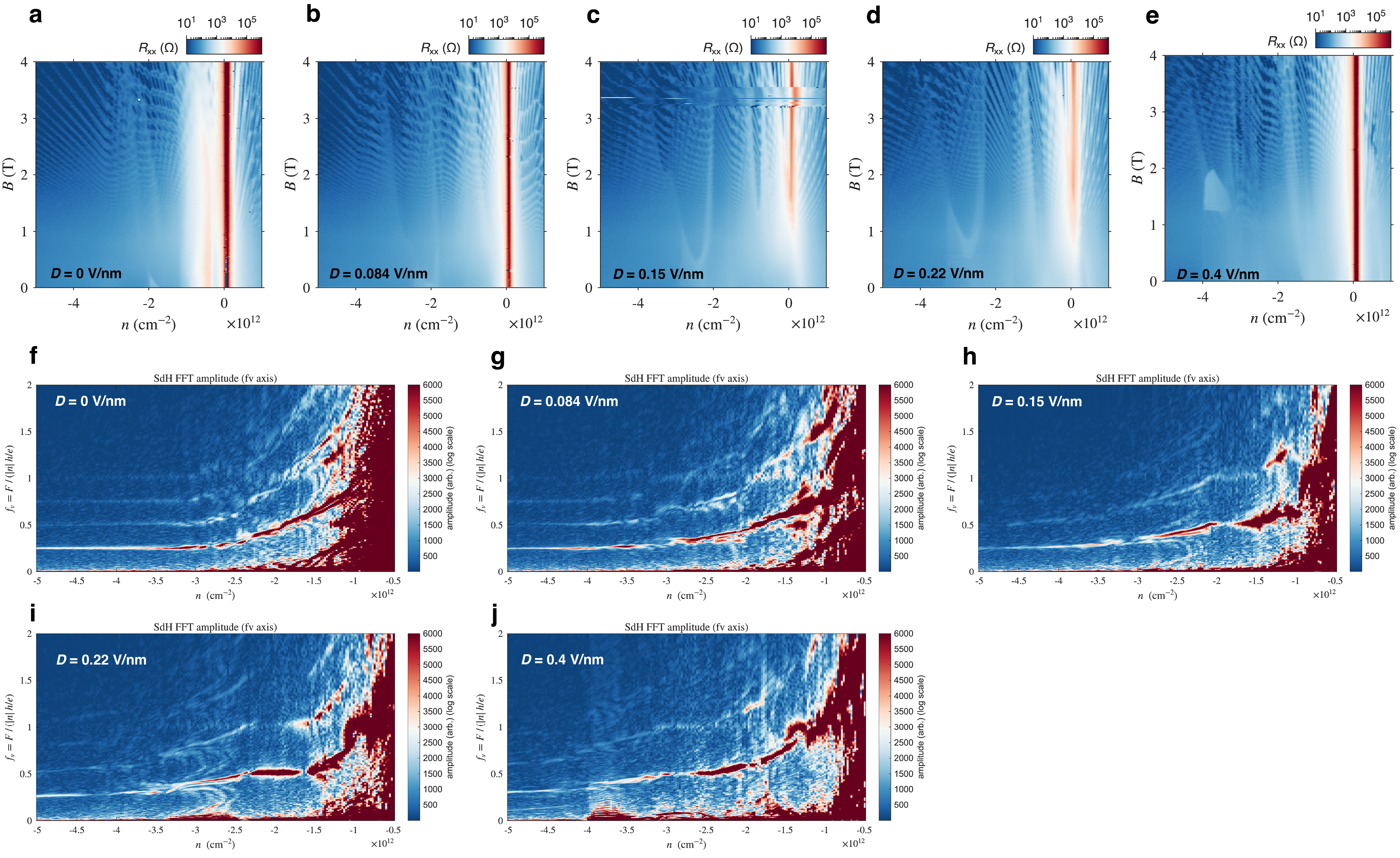} 
    \caption{\textbf{a--e,} Longitudinal resistance $R_{xx}$ as a function of carrier density $n$ and out-of-plane magnetic field $B$ at different fixed $D/\varepsilon_0$ at $T \approx 150\,\mathrm{mK}$. \textbf{f--j,} Corresponding normalized quantum oscillation frequencies \(f_v=F_\text{FFT}/(\lvert n \rvert h/e)\).}
    \label{extfig2}
\end{figure*}

\begin{figure*}[h]
    \centering
    \includegraphics[width=1\linewidth]{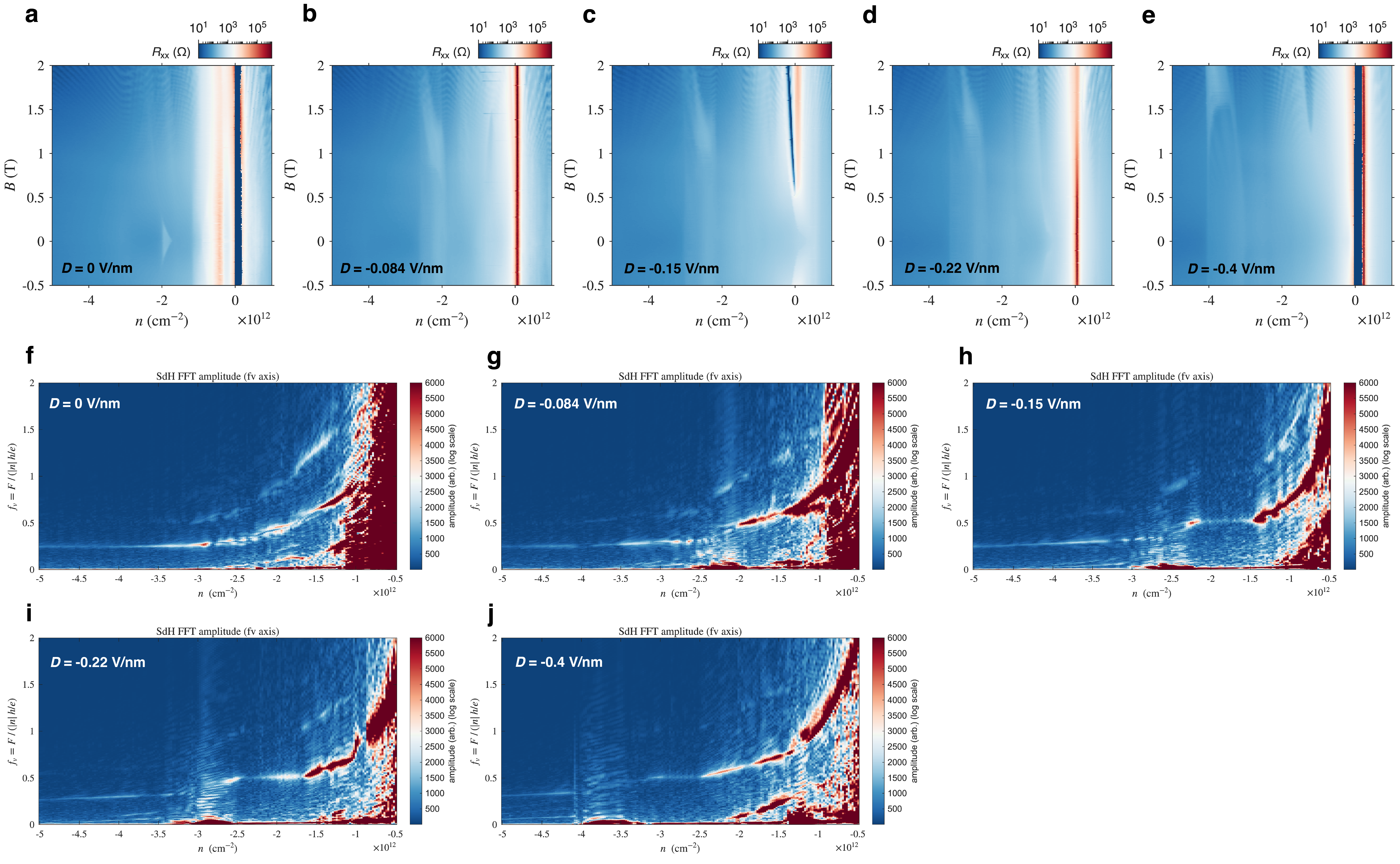} 
    \caption{\textbf{a--e,} Longitudinal resistance $R_{xx}$ as a function of carrier density $n$ and out-of-plane magnetic field $B$ at different fixed $D/\varepsilon_0$ at $T \approx 150\,\mathrm{mK}$. \textbf{f--j,} Corresponding normalized quantum oscillation frequencies \(f_v=F_\text{FFT}/(\lvert n \rvert h/e)\).}
    \label{extfig3}
\end{figure*}

\end{document}